\documentclass[12pt]{article}
\usepackage{pic03}
\usepackage{hyperref}
\usepackage{url}
\usepackage{graphicx}

\usepackage{epsfig}

\newcommand{\LEPI}  {\mbox{LEP1}}
\newcommand{\LEPII} {\mbox{LEP2}}

\newcommand{\RunII} {\mbox{Run II}}

\newcommand{\kt}{\ensuremath{k_{\perp}}}
\newcommand{\WW}{\ensuremath{\mathrm{W}^+\mathrm{W}^-}}
\newcommand{\epem}{\ensuremath{\mathrm{e}^+\mathrm{e}^-}}
\newcommand{\eetoWW}{\ensuremath{\epem\rightarrow\WW}}

\newcommand{\Mw}{\ensuremath{m_{\mathrm{W}}}}
\newcommand{\Gw}{\ensuremath{\Gamma_{\mathrm{W}}}}

\newcommand{\Mt}{\ensuremath{m_{\mathrm{t}}}}

\newcommand{\ppbar}{\ensuremath{\mathrm{p}\bar{\mathrm{p}}}}

\newcommand{\Zz}       {\ensuremath{{\mathrm{Z}^0}}}

\newcommand{\Zqq}      {\ensuremath{\Zz\rightarrow\qq}}

\newcommand{\nuell} {\ensuremath{\nu_{\ell}}}

\newcommand{\lpmnu} {\ensuremath{\ell^{\pm}\nuell}}
\newcommand{\qq}    {\ensuremath{\mathrm{q\overline{q}}}}

\newcommand{\lnln}{\ensuremath{\ell^+{\nu}_{\ell}\ell^{\prime -}\overline{\nu}_{\ell^\prime}}}

\newcommand{\qqln}{\ensuremath{\qq\lpmnu}}
\newcommand{\qqqq}{\ensuremath{\qq\qq}}

\newcommand{\qqlns}{\ensuremath{\mathrm{qq}\ell\nu}}
\newcommand{\qqqqs}{\ensuremath{\mathrm{qqqq}}}
\newcommand{\WWlnln}{\mbox{\WW$\rightarrow$ \lnln}}
\newcommand{\WWqqln}{\mbox{\WW$\rightarrow$ \qqln}}
\newcommand{\WWqqqq}{\mbox{\WW$\rightarrow$ \qqqq}}

\newcommand{\WWg} {\ensuremath{\mathrm{WW}\gamma}}
\newcommand{\WWgg} {\ensuremath{\mathrm{WW}\gamma\gamma}}
\newcommand{\WWZg} {\ensuremath{\mathrm{WWZ}\gamma}}
\newcommand{\kg}{\ensuremath{\kappa_\gamma}}

\newcommand{\etal}    {\mbox{{\it et al.}}}
\newcommand{\Journal}[4] {{#1} \textbf{#2} {#4} (#3)}

\newcommand{\ZP}  {Z. Phys.}
\newcommand{\EPJ} {Eur. Phys. J.}

\begin{document}

\title{\bf W physics at LEP}
\author{ 
  Elisabetta Barberio \\
{\em Physics Department, Southern Methodist University, Dallas, USA }}
\maketitle

\baselineskip=14.5pt
\begin{abstract}
Studying the properties of the W boson plays a key role in
precision tests of the Standard Model. The key measurements performed
over the last decade will be reviewed.
W-pair and
  single-W cross-sections and W decay branching fractions are
  determined and agree well with theoretical predictions.  Including
  the analysis of differential distributions, trilinear and quartic
  couplings of the W boson to the other gauge bosons are extracted.
  The trilinear, C and P conserving couplings are found to be
  $\kappa_\gamma=0.943 \pm 0.055$,
  $\lambda_\gamma=-0.020 \pm 0.024$, and
  $g_1^{\mathrm{Z}}=0.998^{+0.023}_{-0.025}$, consistent with the
  Standard Model expectations. A precise measurement of the mass and
  width of the W boson yields $M_{\mathrm{W}}=80.412 \pm 0.042 \rm GeV$
  and $\Gamma_{\mathrm{W}}= 2.150 \pm 0.091 \rm GeV$. The W mass is in
  good agreement with the one obtained indirectly from an analysis of
  other electroweak data measured at LEP and SLD. Some of the results
  presented in this article are preliminary.

\end{abstract}
\newpage
\baselineskip=17pt

%
%
%
\section{Introduction}

One of the main motivations for the second phase of the LEP $\rm e^+e^-$
storage ring at CERN (LEPII) is the study of the W properties and 
production for a thorough test of the Standard Model (SM) of electroweak
interactions \cite{nobel}.
Between 1996 and 2000, the LEP collider  was operated
at centre-of-mass energies above the \WW\ production threshold,
allowing for investigation of different aspects 
of W-pair production in  $\rm e^+e^-$ annihilation, which are
crucial test of the Standard Model of electroweak interactions.
 The four LEP collaborations 
(ALEPH, DELPHI, L3, and OPAL) collected
a total sample of around 40,000 W pairs in the 2.8~$fb^{-1}$ of data
recorded. 

\section{W-pair and single-W production}

The process \eetoWW\ can be identified with high efficiency
in all decay modes of the W boson. 
\WW\ events are classified into three final states, according to their
decay modes.  \WWqqqq\ events
comprise 45\% of the total \WW\ cross-section and are characterised
by four energetic jets of hadrons with little or no missing energy.
Semi-leptonic \WWqqln\ decays comprise 44\% of the total \WW\
cross-section and are characterised by two distinct hadronic jets, a
high-momentum lepton and missing momentum due to the prompt neutrino
from the leptonic W decay.  
The $\WWlnln$ channel events have at least two unobserved neutrinos
and a relatively low branching fraction, 11\% .

\begin{figure}
\begin{center}
 \epsfig{figure=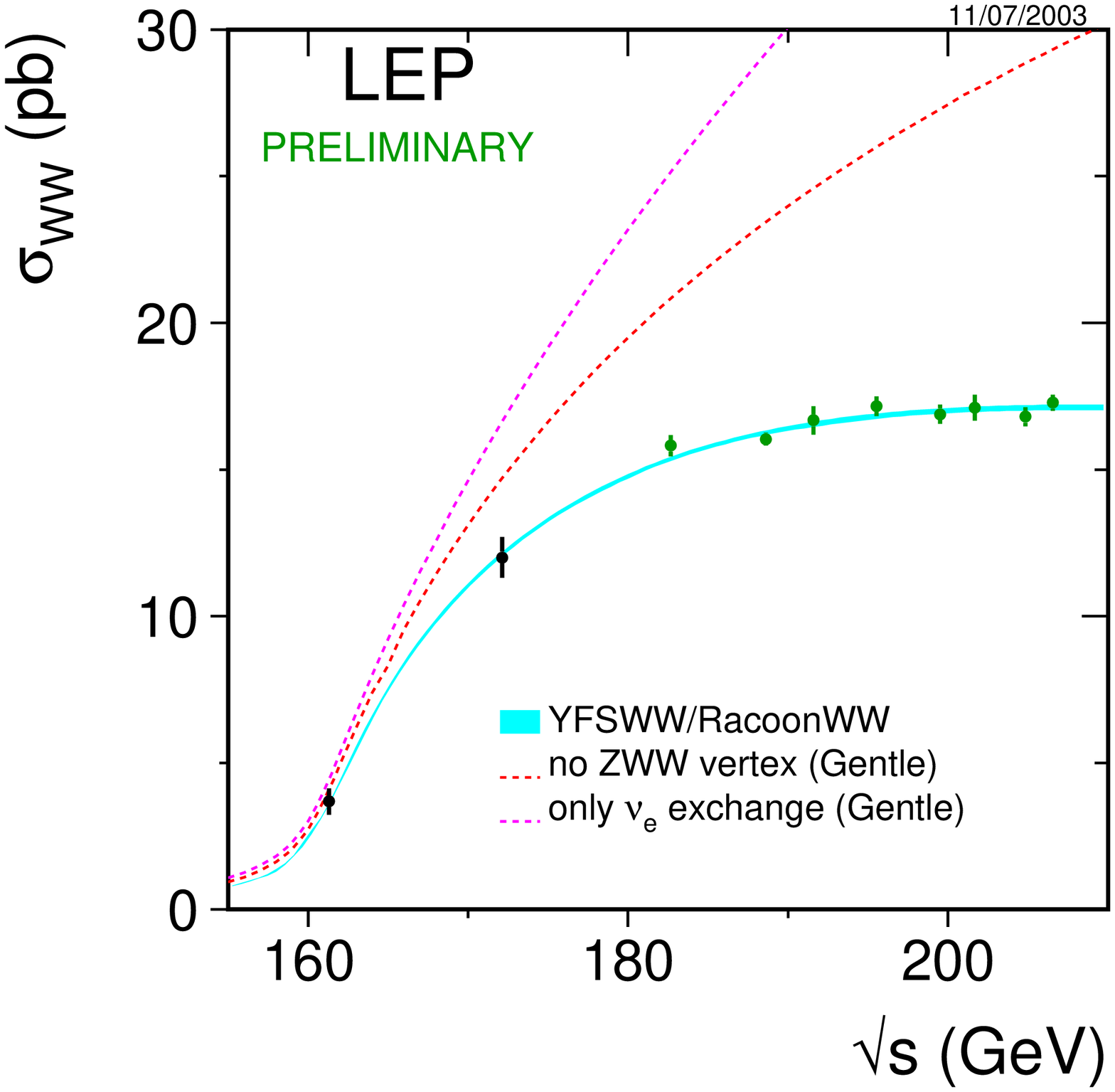,width=5.0cm}
 \epsfig{figure=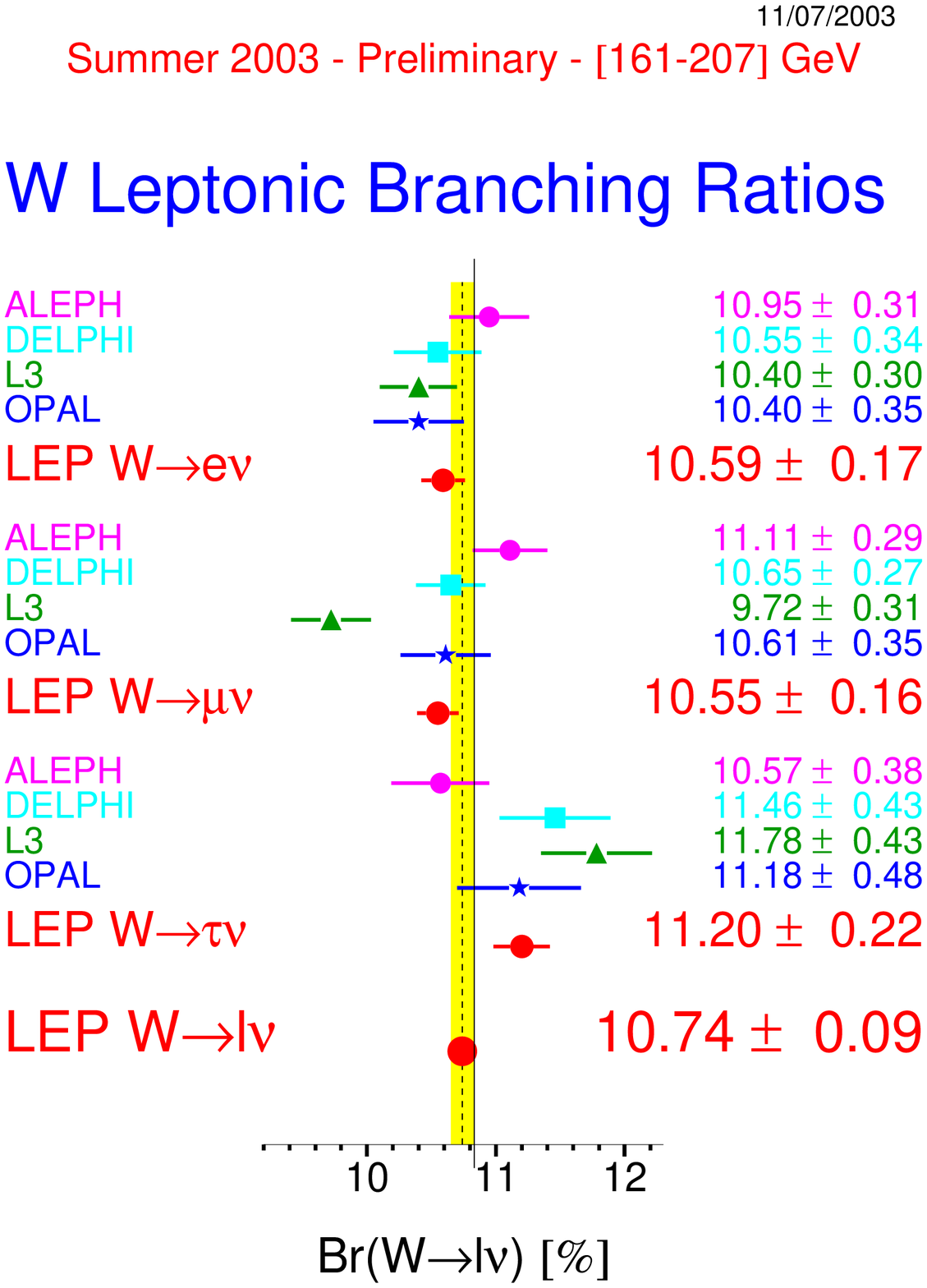,width=4.5cm}
\end{center}
\vspace{-1.0cm}
\caption{
 Measurements of the W-pair production cross-section and W leptonic 
 branching fractions. The measured e W-pair production cross-section
 is compared to theoretical predictions (curve). 
\label{sww_vs_sqrts}}
\end{figure}

At LEP W bosons are produced in pairs in the process $eetoWW$.
About 95\% of the resonant W-pair production 
is described in the SM by three charged current Feynman
diagrams, one $t$-channel diagram with neutrino exchange and two
$s$-channel diagrams with $\gamma$ and Z exchange.
The total production cross-section, $\sigma_{WW}$, measured by the four LEP 
experiments \cite{cc} at the centre of mass energies
 between 161 and 209 GeV is shown in 
Figure~\ref{sww_vs_sqrts}, where data points
are compared with the theoretical calculations, 
\cite{4f_bib:yfsww}.
The  $\sigma_{WW}$ evolution in case the $\gamma WW$ or additionally 
the $ZWW$ are missing is also shown in 
Figure~ \ref{sww_vs_sqrts}. The measurements clearly indicate
the non-Abelian nature of the SM of electroweak interactions.

Measurement of the W branching fractions is a test of lepton 
universality at high $q^2$. The measured values of the individual
leptonic W branching fraction  support lepton universality,
Figure~\cite{4f_bib:yfsww}, and
can be combined to give $Br(W\to \ell\nu)=(10.74\pm 0.09) \%$ \cite{cc},
  in agreement with the SM expectations.
The W decay rate into quarks pairs
is measured to be $Br(W\to {\rm qq'})=(60.77\pm 0.28) \%$, \cite{cc}, 
 in agreement with the SM expectations.

\section{Triple Gauge coupling}

The non-Abelian structure of the SM predicts the existence of coupling
between the gauge bosons. The existence of the Triple Gauge Coupling (TGC)
 $\gamma WW$ and $ZWW$ is unambiguously confirmed by measurements of 
$\sigma_{WW}$ at LEP2 (see above).
In the most general Lorentz invariant ansatz the  TGC vertices are 
 parametrised by 14 couplings \cite{tgc-theory}. Imposing C and P invariance 
and SU(2) symmetry, only three couplings are left to be studied:
$\lambda_\gamma$, $k_\gamma$ and $g_1^Z$. In the SM the values of 
these couplings are:
$\lambda_\gamma=0,~ k_\gamma =1,~ g_1^Z=1 .$
$g_1^Z$
describes the $ZWW$ vertex, while $\lambda_\gamma$ and $k_\gamma$ 
are related to the static magnetic dipole and electric 
quadrupole of the W boson.

TGCs affect the total production cross-section, the shape of
the differential cross-section as a function of the polar W$^-$
production angle and the  polarization of the W. 
Deviations from the SM would lead to a modification
of these quantities.
Studies of the TGCs are performed by 
exploiting the information contained in the differential 
distributions of W boson production ($\theta_W$) and decay angles. 
  The analyses presented by each
experiment make use of different combinations of each of these
quantities.  In general, however, all analyses use at least the
expected variations of the total production cross-section and the
W$^-$ production angle. 
The measured multi-differential cross-section is compared to theoretical 
expectation, for which it is as important as for the total 
cross-section to take high-order electroweak corrections into account.
At LEP additional information on TGCs can be obtained from single W 
production, which is particularly sensitive to \kg.
Hence,  some experiments include this channel in their
analyses.

\begin{figure}
\begin{center}
 \epsfig{figure=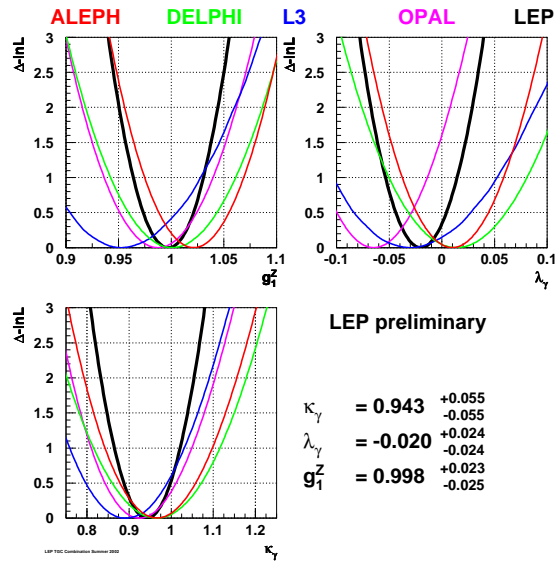,width=8.0cm}
\end{center}
\vspace{-1.0cm}
\caption{
Likelihood curves for $\lambda_\gamma$, $k_\gamma$ and $g_1^Z$
for the four LEP experiments and the LEP combination.
\label{tgc1}}
\end{figure}
The likelihood curves for $\lambda_\gamma$, $k_\gamma$ and $g_1^Z$
for the four LEP experiments \cite{tgc} are shown in Figure~\ref{tgc1}. 
The combination of
these measurement yields, \cite{tgc}: 
$$  k_\gamma=0.943 \pm 0.055,
~ \lambda_\gamma =-0.020 \pm0.024, 
~ g_1^Z=0.998^{+0.023}_{+0.025},$$
a 5\% precision measurement in good agreement with  SM prediction.
The dominant error is from higher electroweak corrections.

\section{W polarization}
In addition to the two possible transverse polarization states of massless
spin-1 particle (e.g.$\gamma$), massive gauge boson should exist also 
in the longitudinal state. In the SM the polarization state is related 
to the mechanism of electroweak symmetry breaking in which three 
degrees of freedom of the scalar Higgs field generate the longitudinal
states of the W and Z. Only transverse 
bosons are produced in weak process involving light fermions, while 
a considerable contribution from longitudinal prolarized 
W is expected in in W-pair production at LEP, making this helicity
state experimentally accessible.

\begin{figure}
\begin{center}
 \epsfig{figure=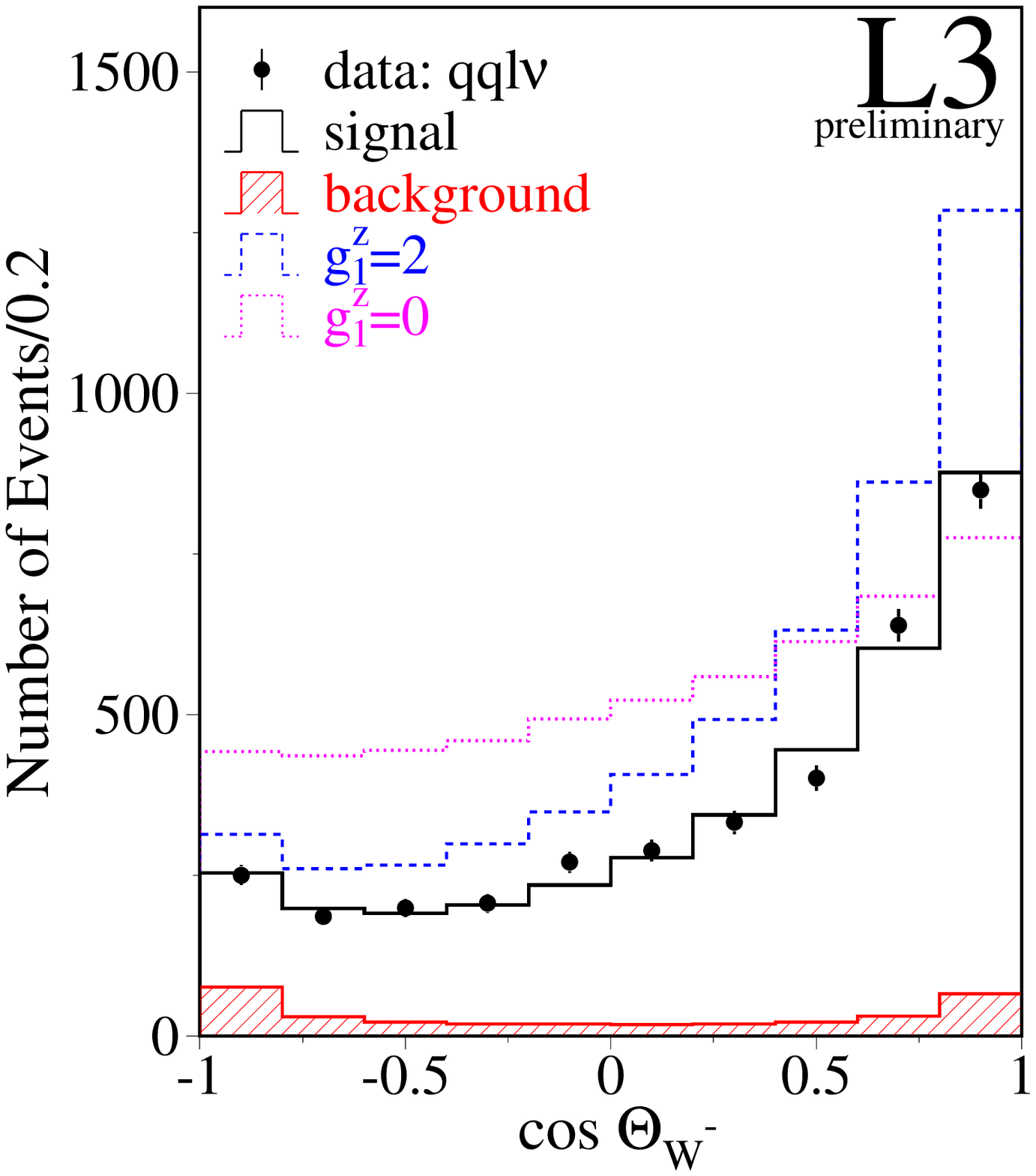,width=4.0cm} \\
 \epsfig{figure=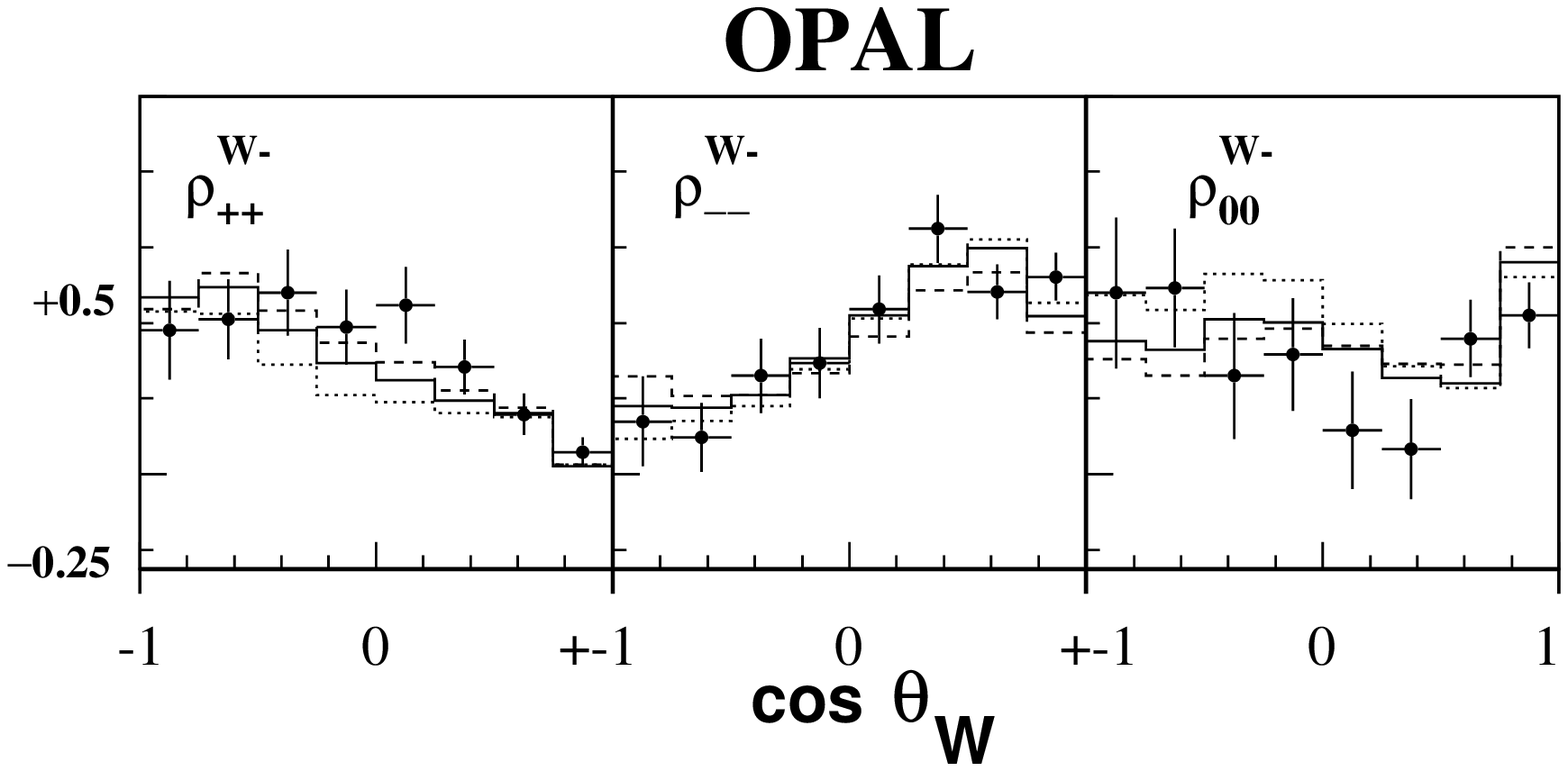,width=9.0cm}
\end{center}
\vspace{-1.1cm}
\caption{At the top, the L3 polar angular distribution.
At the bottom, the OPAL SDM distribution.
\label{pola}}
\end{figure}

L3 extracted the different helicity states of the W boson by exploiting
the angular distribution of the W decay products in the W rest frame.
They fitted the expected angular distribution for the different 
helicity states to the data, corrected for efficiency and background
The result clearly establish the existence of the longitudinal helicity
state, Figure~\ref{pola}.
OPAL  measured the W polarization using the Spin Density Matrix (SDM) 
method. The 
polarised differential cross sections are derived by multiplying 
the measured differential cross section with the corresponding diagonal 
elements of the single particle SDM element, corrected 
for efficiency and background. The polarization is then 
obtained by integrating over $\cos \theta_W$. The OPAL result
also clearly show the existence of the longitudinal helicity state, 
Figure~\ref{pola}.

\section{Quartic Gauge coupling}
Within the Standard Model, quartic electroweak gauge boson vertices
with at least two charged gauge bosons exist. In $e^+e^-$ collisions at
\LEPII\ centre-of-mass energies, \WWZg\ and \WWgg\ vertices
contribute to \WWg\ production.
However,  their existence cannot be proven because at LEP energies
the effect of SM quartic coupling is too small to be measurable. Hence,
only limits on anomalous contributions to the quartic vertices are
derived.
Recently, OPAL performed an analysis in the \WWg\ final state
\cite{mark}.
The measured photon rate and spectrum is in agreement with the SM 
calculations. These data 
 are used to derive
95~\% confidence level upper limits
on possible anomalous contributions to the
 \WWZg\ and \WWgg\ vertices:
\begin{eqnarray*}
   -0.020~\mathrm{GeV}^{-2} < & a_0/ \Lambda^2 & < 0.020~\mathrm{GeV}^{-2}, \\
   -0.053~\mathrm{GeV}^{-2} < & a_c/ \Lambda^2 & < 0.037~\mathrm{GeV}^{-2}, \\
   -0.16~\mathrm{GeV}^{-2} < & a_n/ \Lambda^2 & < 0.15~\mathrm{GeV}^{-2},
\end{eqnarray*}
where $\Lambda$ represents the energy scale for new physics and $a_0$,
$a_c$ and $a_n$ are dimensionless coupling constants.

\section{W boson mass}
At tree level, the electroweak observables are fully determined
by the mass of the Z boson, the Fermi constant,
the electromagnetic coupling and the CKM matrix elements.
Due to higher-order radiative corrections,
the simple tree
-level predictions are modified such 
Standard Model observables depend also on the strong coupling constant, 
the top mass and to lesser extent to the Higgs mass.
In this context, the mass of the W boson provides indirect 
knowledge on the Higgs mass through 
higher-order radiative corrections ($\log(m_H^2/m_Z^2)$) and 
its precise measurement allows predictions of the Higgs mass.
The final goal of \LEPII\ is to measure \Mw\ with a precision 
of about 30-35 MeV; a lower uncertainty 
will not improve the knowledge on the Higgs mass, as the limiting factor is 
the current experimental precision of the top mass.

The first precision measurements of the W-boson mass 
were performed at \ppbar\ colliders. 
Using a W sample exceeding 200,000 events, 
CDF and D0 combined achieved:
$\Mw = 80.454 \pm 0.060 \rm ~GeV$,  \cite{bib:tevmass}.
The sample of W bosons collected at \LEPII\ is significantly
smaller, but the mass measurement benefits from a
clean environment which allows more information to be
extracted from each recorded event.

At an $e^+e^-$ collider  \Mw\ 
can be either  derived from the \WW\ threshold cross section or 
from he direct reconstruction of the W boson's invariant 
mass from the observed W decay products on an event-by-event basis.
For most of the time 
\LEPII\ has operated at energies significantly above the $\WW$ 
threshold, where the $\epem\rightarrow\WW$ cross section has little 
sensitivity to $\Mw$. Hence, only the direct 
reconstruction method is discussed here.
Also the $\WWlnln$ channel has limited \Mw\ sensitivity
and is not discussed.

The invariant masses of the two W bosons are determined directly from
the reconstructed momenta of observed decay products. 
Hadrons are grouped together into jets using clustering algorithms
such as \kt.  In \qqlns\ events, charged leptons are identified and
neutrinos are inferred from the missing energy and momentum. 

Experimentally, the limiting factor in the mass resolution is the 
uncertainty in the jet energy measurement, which is poor in contrast 
to the measured jet directions.  As the centre-of-mass energy is well 
known, the mass resolution can be improved significantly 
(factor $\sim$2--3) by imposing the constraints of
energy and momentum conservation.

\begin{figure}[htb]
\begin{center}
   \epsfig{file=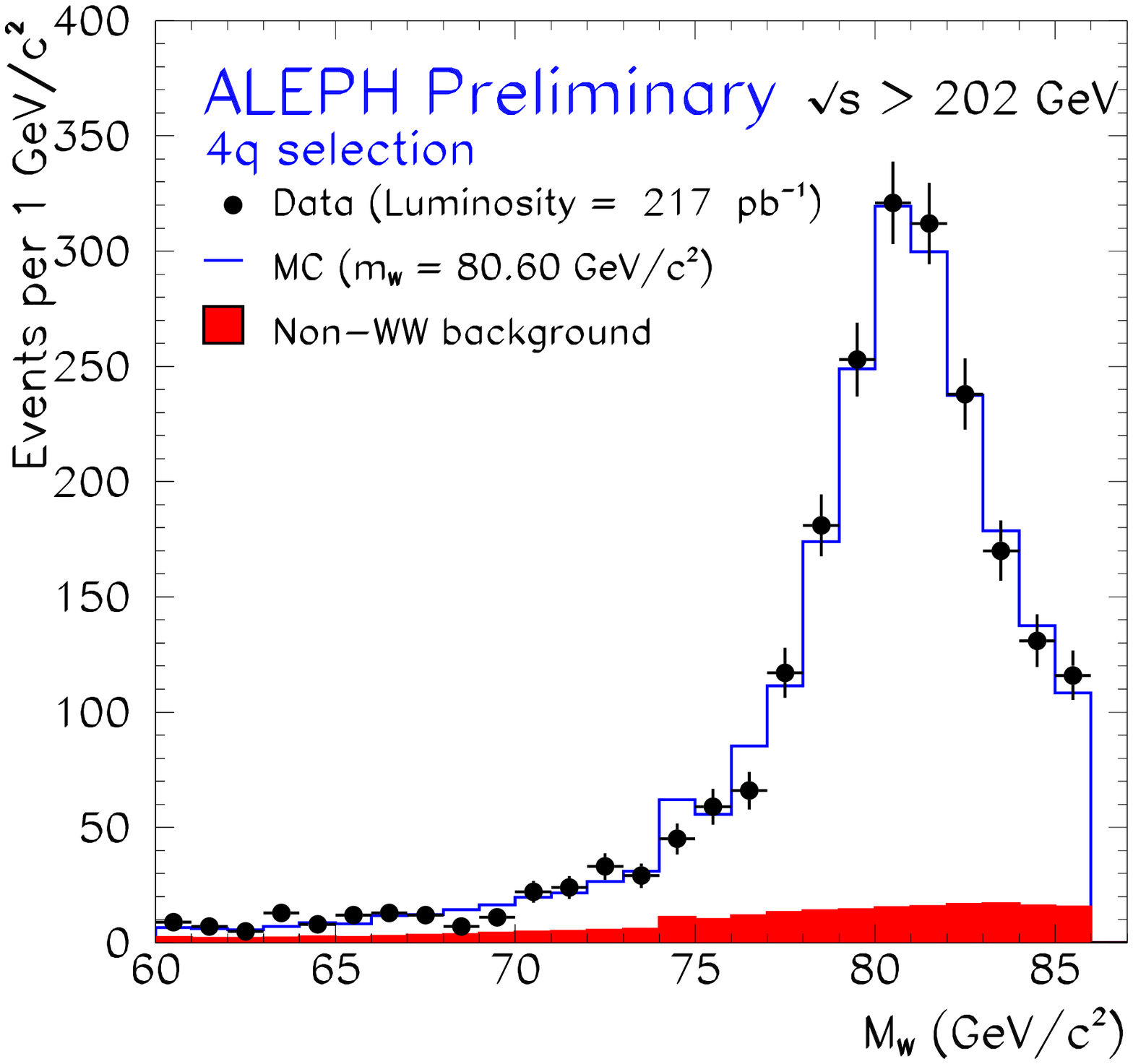,width=3.7cm}
   \epsfig{file=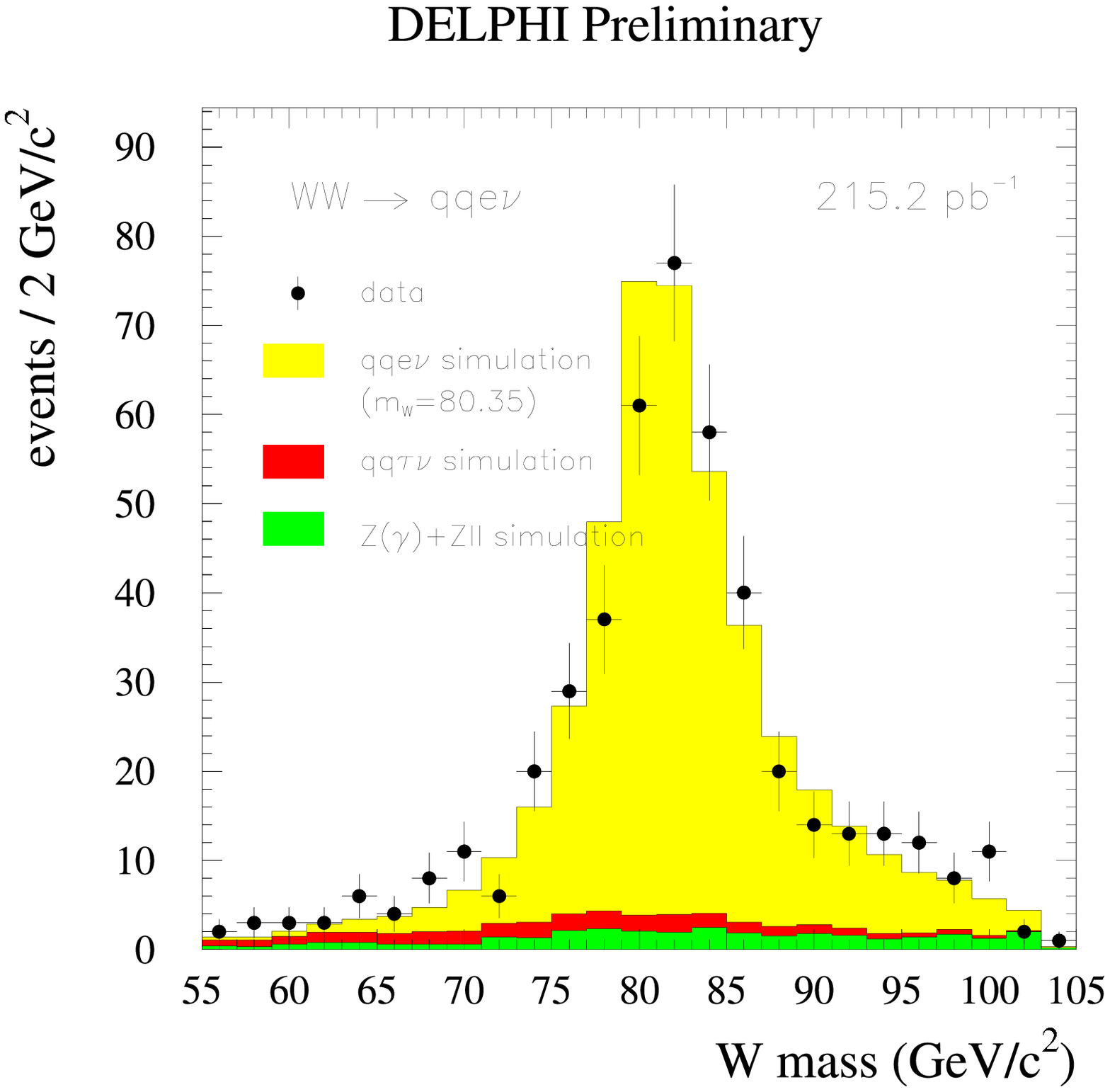,width=3.7cm} \\
   \epsfig{file=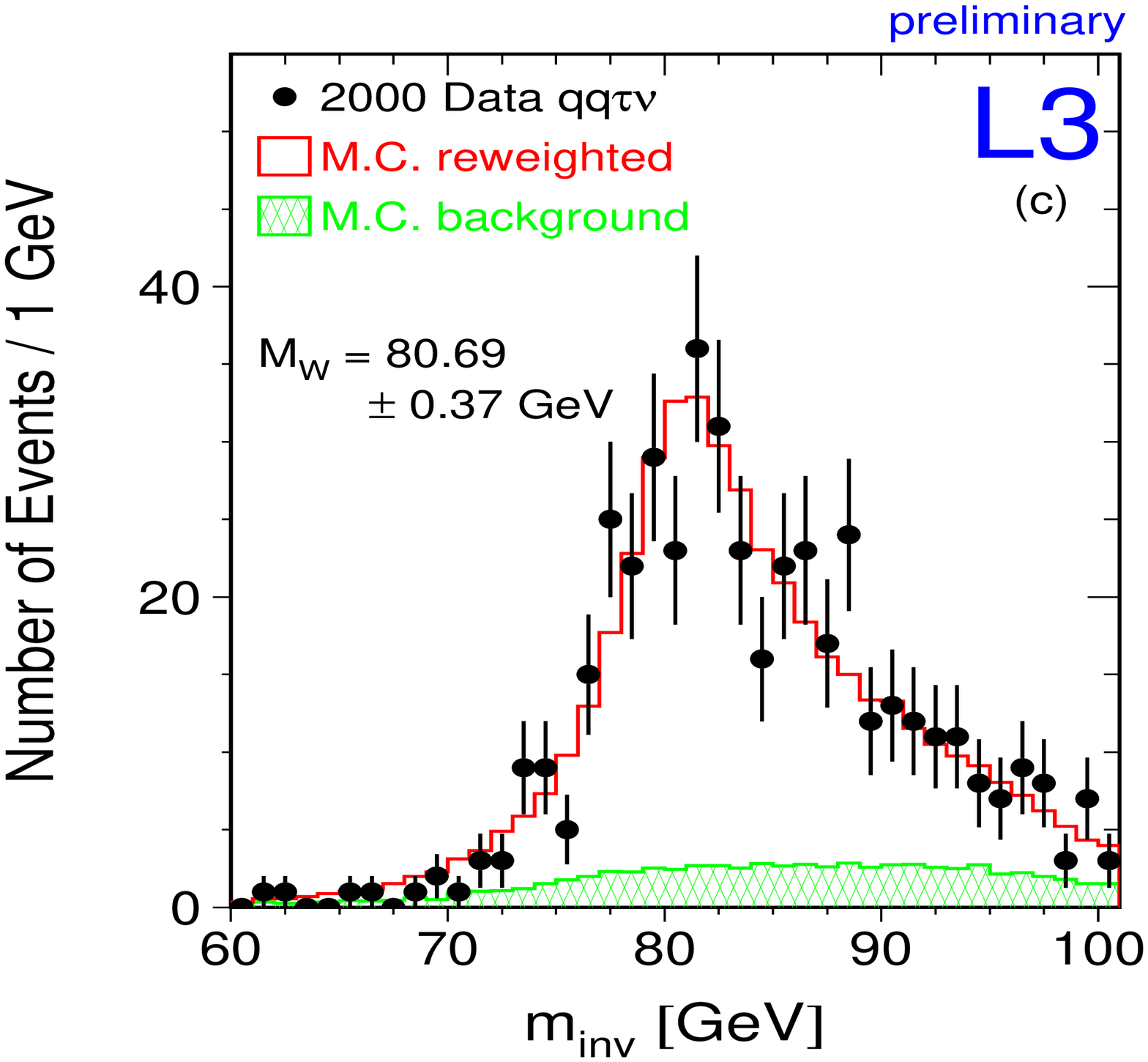,width=3.7cm}
   \epsfig{file=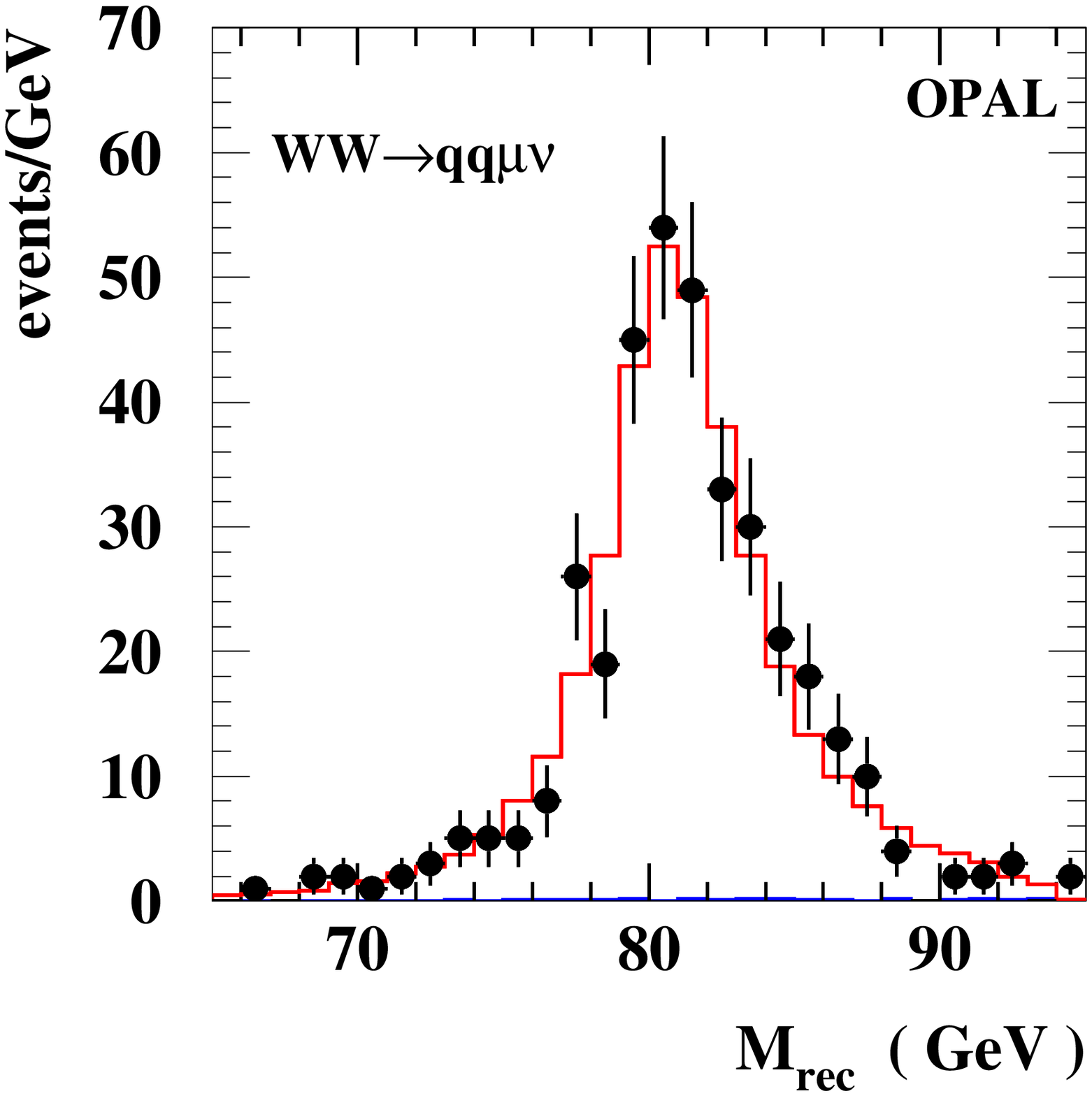,width=3.7cm}
    \vspace{-1.cm}
\end{center}
    \caption{
      Reconstructed mass spectra from the LEP collaborations.
      \label{fig:mass}
      }
\end{figure}

The reconstructed mass spectrum looks rather different from a pure
relativistic Breit--Wigner distribution for several reasons.
For example, the presence of initial state radiation (ISR) means
that the energy producing the W pairs is always less than
twice the incoming beam energy.
This causes a tail toward higher invariant masses, as
the collision energy assumed in the kinematic fit is overestimated.
The detector resolution also tend to significantly
smear out the line-shape as the experimental resolution is not
significantly better than the W boson width for most channels.
As a result, the W boson mass cannot be extracted by simply 
fitting an analytic Breit--Wigner shape, but all 
extraction methods need to be calibrated 
on a Monte Carlo simulation which includes all the various effects to model
the dependence of the spectrum on \Mw. 
Most of the systematic errors associated to \Mw\ account for
effects which may be missing in this Monte Carlo.
Figure~\ref{fig:mass} shows some reconstructed mass spectra compared to
the Monte Carlo predictions.

The most important uncertainties for the LEP W mass measurement are
related to the modeling of the detector response, modeling
the hadronization of quarks into jets, understanding the LEP beam
energy, and final state interactions.
To limit the uncertainty from detector modeling, the LEP collider
was run from time to time throughout each running year at the
\Zz\ resonance to take large samples of \Zz\ decays.
These data, as well as similar samples at high energy, are 
used to study the detector response to leptons and jets and limit 
the deficiencies in the detector models.

Hadronization uncertainties are estimated by comparing different
Monte Carlo implementations of the hadronization process and
 re-weighting key variables in
 Monte Carlo to correspond to data and propagating the effect to \Mw.

The relative uncertainty in the LEP beam energy enters directly
 into the uncertainty in \Mw, due to the use of kinematic fits.
 Uncertainties in
 beam energy are taken from the extrapolation to
 high energy the result of the resonant
 depolarisation measurement, which can only be
 done for beam energies below 60 GeV \cite{bib:lepenergy}.

 A significant bias in the apparent W mass measured in the \qqqq\
 channel could arise if the hadronisation of the two W bosons is not
 independent and correctly modeled. 
 Standard Monte Carlo models assume that the two systems decay
 independently. However, interactions that can exchange momentum between the
  two W systems could distort the final W line-shape.
Two specific phenomena  are known to exist, but with rather
  uncertain strengths: colour reconnection (CR) \cite{bib:CR} 
and Bose--Einstein
correlations (BEC) \cite{bib:BE}.
 Effects on the mass of such interactions are estimated by using
 phenomenological models and direct searches for these effects limit 
the viable set of such models.

The four LEP collaborations combine their results taking into
account systematic uncertainties which are correlated between
channels, experiments, and years of LEP running.
This combination procedure is still evolving, as better
information about the nature of these various correlations
becomes available.

At present, the preliminary combined LEP W mass result from direct 
reconstruction is \cite{bib:m01-lepmw}:
$\Mw = 80.412\pm0.042,\mathrm{GeV}.$
As can be seen in the detailed breakdown of the direct measurements 
uncertainties shown 
in Table~\ref{errors}, the \qqqqs\ channel has rather large
uncertainties associated with Bose--Einstein correlations and colour
reconnection.
Due to these uncertainties, the \qqqqs\ channel carries a weight
of only 9\% in the combined result, even though it is statistically
more precise.
\begin{table}[tbp]
 \begin{center}
  \begin{tabular}{|l|r|r||r|}\hline
       Source  &  \multicolumn{3}{|c|}{ $\delta \Mw$ (MeV)}  \\
                             &  \qqln & \qqqq  & combined  \\ \hline
 ISR/FSR                     &  8 &  8 &  8 \\
 Hadronisation               & 19 & 18 & 18 \\
 Detector Systematics        & 14 & 10 & 14 \\
 LEP Beam Energy             & 17 & 17 & 17 \\
 Colour Reconnection         & -- & 90 &  9 \\
 Bose-Einstein Correlations  & -- & 35 &  3 \\
 Other                       &  4 &  5 &  4 \\ \hline
 Total Systematic            & 31 & 101 & 31 \\ \hline
 Statistical                 & 32 & 35  & 29 \\ \hline
 Total                       & 44 & 107 & 43 \\ \hline
 \end{tabular}
\end{center}
\vspace{-.5cm}
\caption{Error decomposition for the combined LEP W mass results.}
 \label{errors}
\end{table}

The direct reconstruction method employed at \LEPII\ is
sensitive to the W width as well as the W mass.
In the standard mass analyses, the width i fixed
to the Standard Model expectation for a given mass value.
The width is extracted allowing it to be a second free parameter
in the fits. The LEP combined value of measurements is 
$ \Gw = 2.150 \pm 0.091~ \rm GeV$ 
in agreement with the the SM expectation.

\subsection{Colour Reconnection}

 In \qqqqs\ events, the products of the W decays in general have a
 significant space-time overlap as the separation of their decay
 vertices is small compared to characteristic hadronic distance
 scales.  Colour reconnection refers to a rearrangement of the colour
 flow between the two W bosons.  The effects of interactions between
 the colour singlets during the perturbative phase are expected to be
 small.
 The situation is less clear in the non-perturbative phase, where
 phenomenological models are implemented in hadronic Monte Carlos.  A
 higher susceptibility to CR (and more \Zqq\ background) is expected
 when $\rm W^+$ and $\rm W^-$ hadronisation regions overlap, so the space-time
 picture of the QCD shower development is important.

 The predicted (barely) observable effects of CR include changes to the
 charged particle multiplicity, momentum distributions and the
  particle flow relative to the 4-jet topology.  
 
Colour reconnection effects tend to enhance or suppress particle
production in the regions between the main jets.
Currently, all four LEP collaborations are pursuing analyses
aimed at measuring the particle flow distribution in \qqqqs\
final states with the ultimate aim of discriminating between
various CR models.
Combining the results of these analyses, taking into account
their different sensitivities, it is found that the no-CR 
 scenario agrees with data only at the level of 2$\sigma$ and a 
moderate reconnection fraction is preferred. However, no definitive
conclusion can be drawn.

The most sensitive estimator of CR is the invariant mass of the W boson
measured in the \qqqqs\ channel. Removing low momentum 
particles reduces the bias due to CR for all investigated models.
 This can be achieved in two ways, either with a cut on 
the particle momentum either using a modified
jet algorithm. Both methods give similar results.
All LEP experiment are doing this analyses and DELPHI 
already presented some preliminary results in \cite{delphi}.
Since this method is almost uncorrelated with the particle flow method,
a combination of the two is foreseen. 

With all four LEP experiments combined, it is likely that one 
can achieve a 5$\sigma$ evidence of CR and be able to reduce the
significantly the uncertainty on the W mass from its
current value.

\subsection{Bose-Einstein correlation}

Bose-Einstein correlation leads to the enhanced production of
identical boson pairs, such as $\pi_1^+\pi_2^+$ or $\pi_1^-\pi_2^-$, 
at small 4-momentum
difference, $Q^2_{1,2}$.  This phenomena is firmly
established  in hadronic \Zz\ at \LEPI\ and between
the particles of a single W boson at \LEPII.
Since the W boson decay length (0.1~fm) is significantly shorter
than the hadronization scale (1~fm), it is entirely plausible
that there can be additional BE effects between particles originating
from different W bosons in \qqqqs\ events.

Traditionally, BEC is studied using a 2-particle correlation function:
$R_{1,2}=\rho_2(1,2)/\rho_0(1,2),$
where $\rho_2$ and $\rho_0$ are 2-particle densities with and without
BEC, respectively.  
One serious problem in this area is the construction of the reference
sample, $\rho_0$, in a model independent way.
All four LEP collaborations use the technique described in \cite{bib:eddi}
to construct such a sample (apart from 
background subtraction). This method 
involves mixing pairs of data events, such as the
hadronically decaying W in \qqlns\. 
Data from two semi-leptonic \qqlns\
events are mixed (without the lepton) and compared to
data from genuine \qqqqs\ events.
In a rigorously model--independent test, these two samples
should look identical if there is no BEC present between the
decay products of different W bosons. Some of the LEP 
experiments are still finalizing their results.

The systematic uncertainty on \Mw\ due to BEC will be derived 
from the final combined LEP results.  The mass bias is expected 
to be reduced to few MeV.
\begin{figure}
\begin{center}
    \epsfig{file=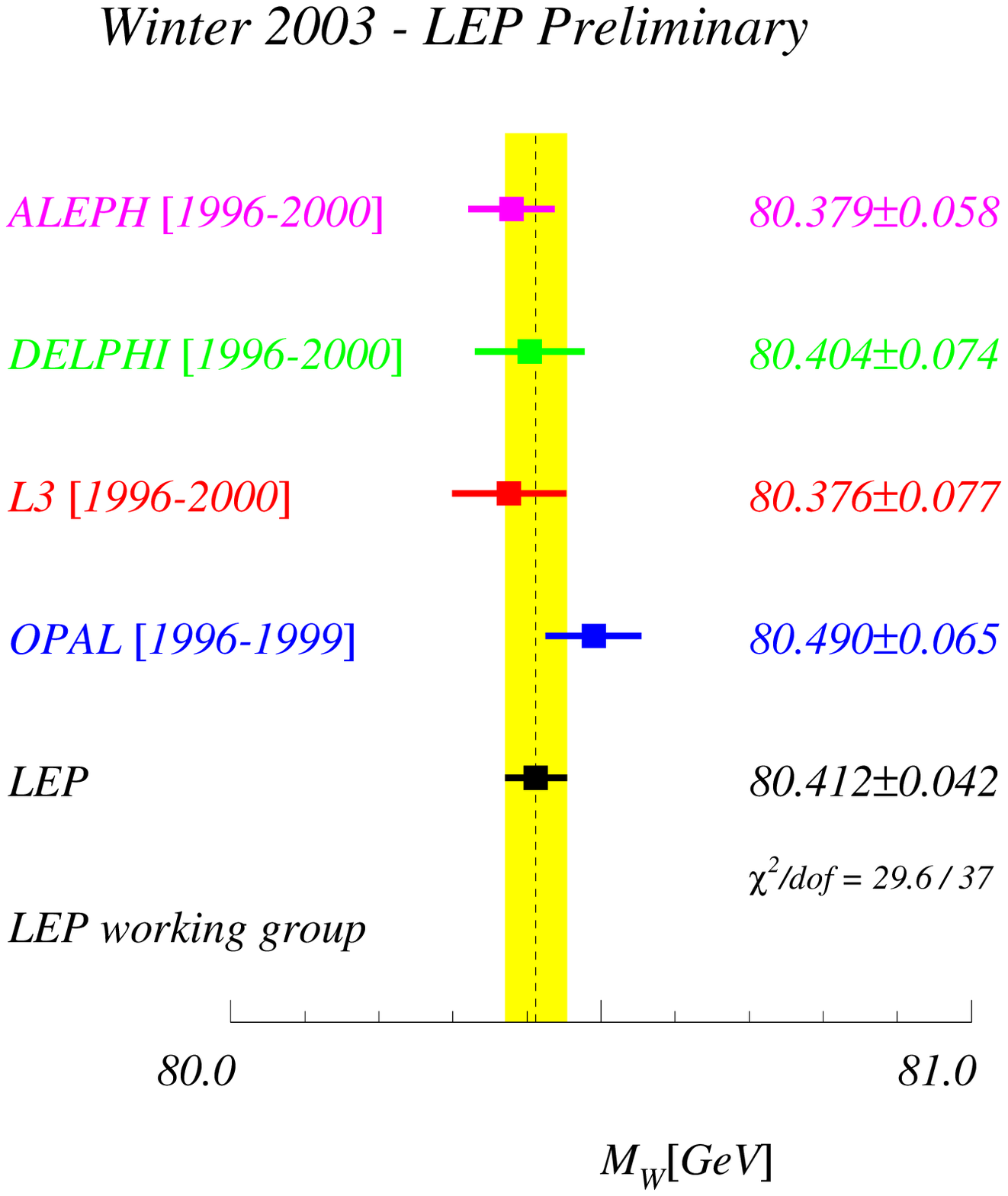,width=4.5cm}
    \epsfig{file=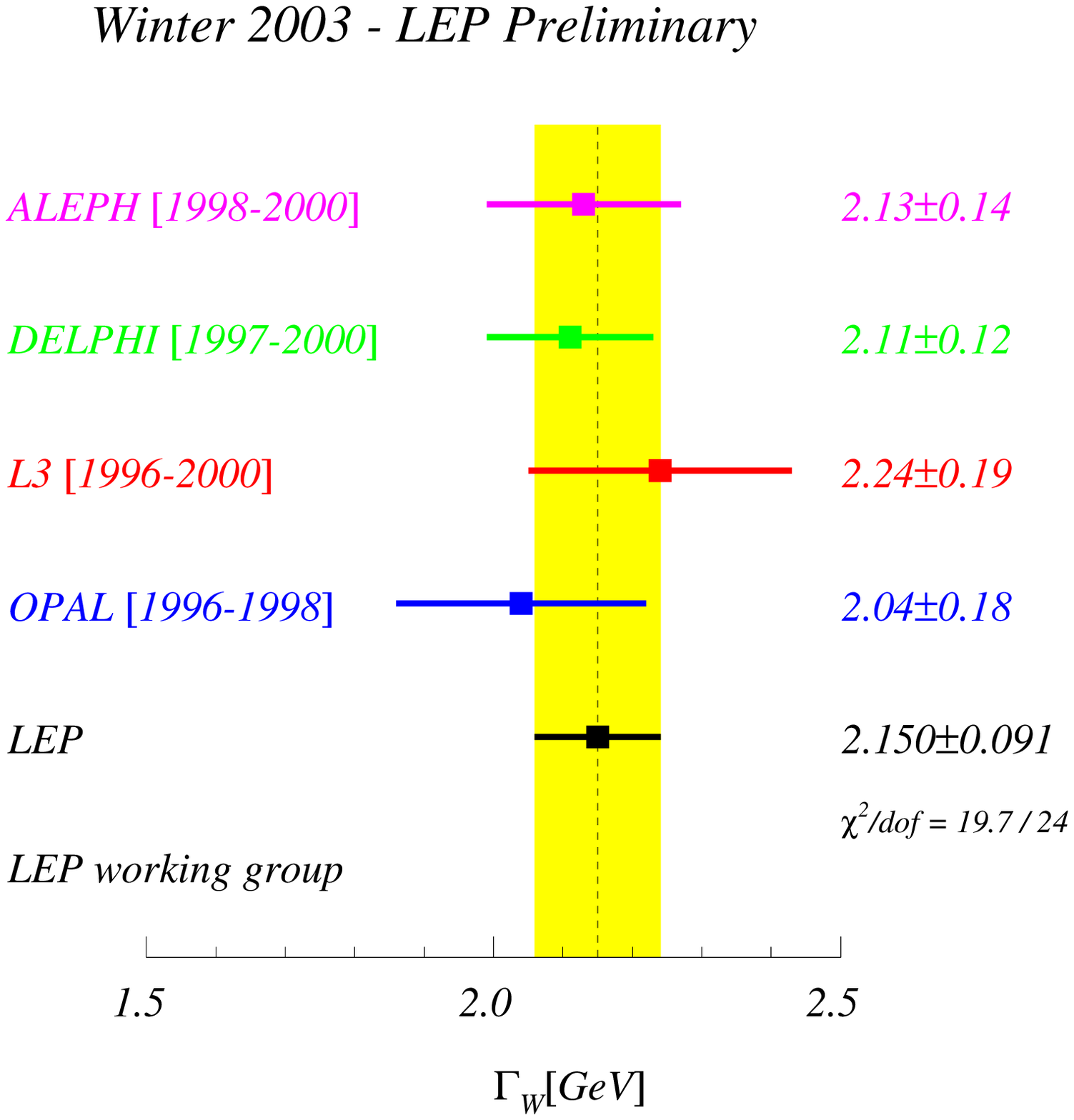,width=4.5cm}
\end{center}
    \vspace{-1.0cm}
    \caption{
      LEP results on the mass and  width of the W boson.
      \label{fig:ab}}
\end{figure}

%
%

\section{Conclusions and Perspectives}
\begin{figure}[htb]
\begin{center}
    \epsfig{file=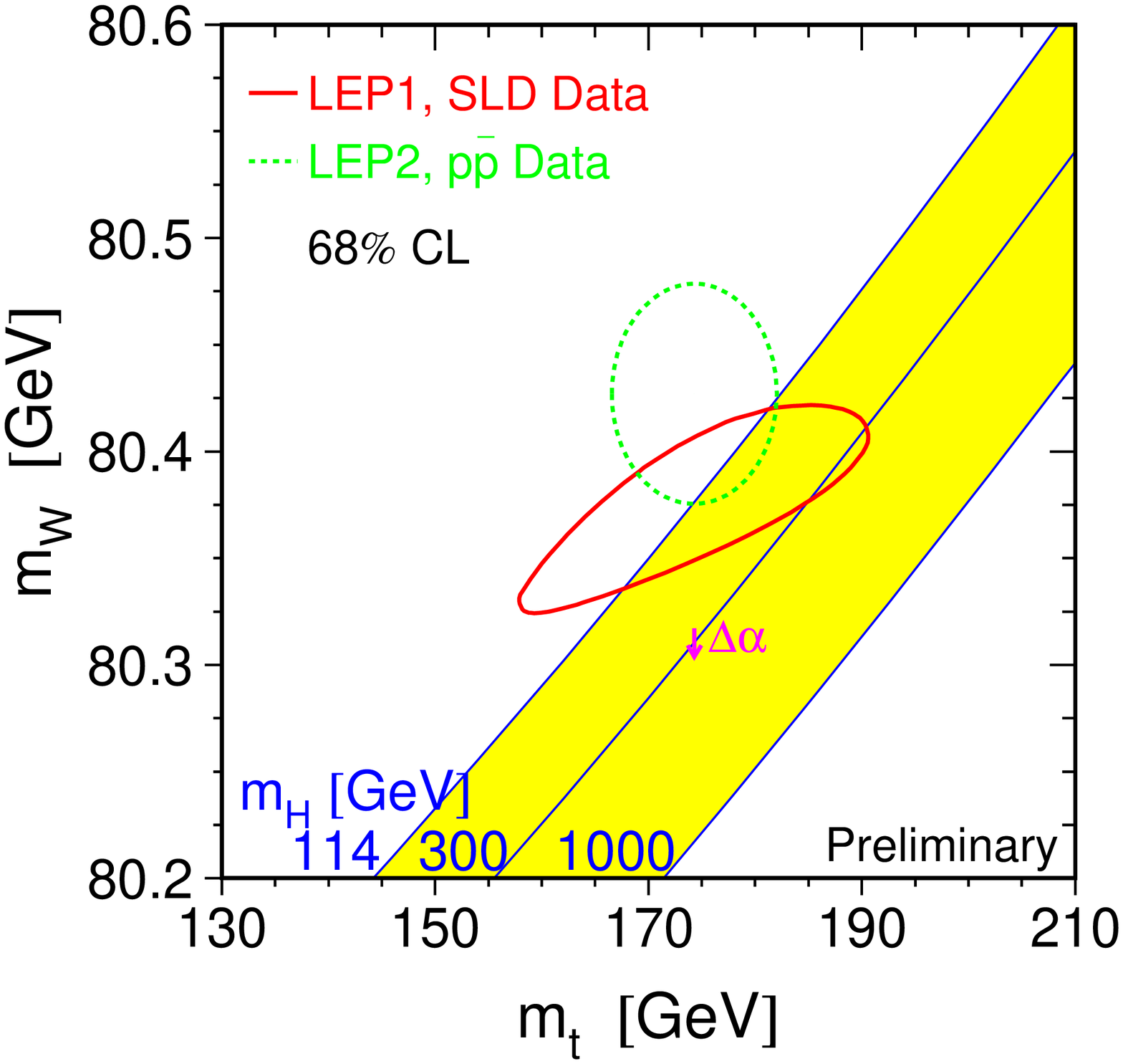,width=4.6cm}
    \epsfig{file=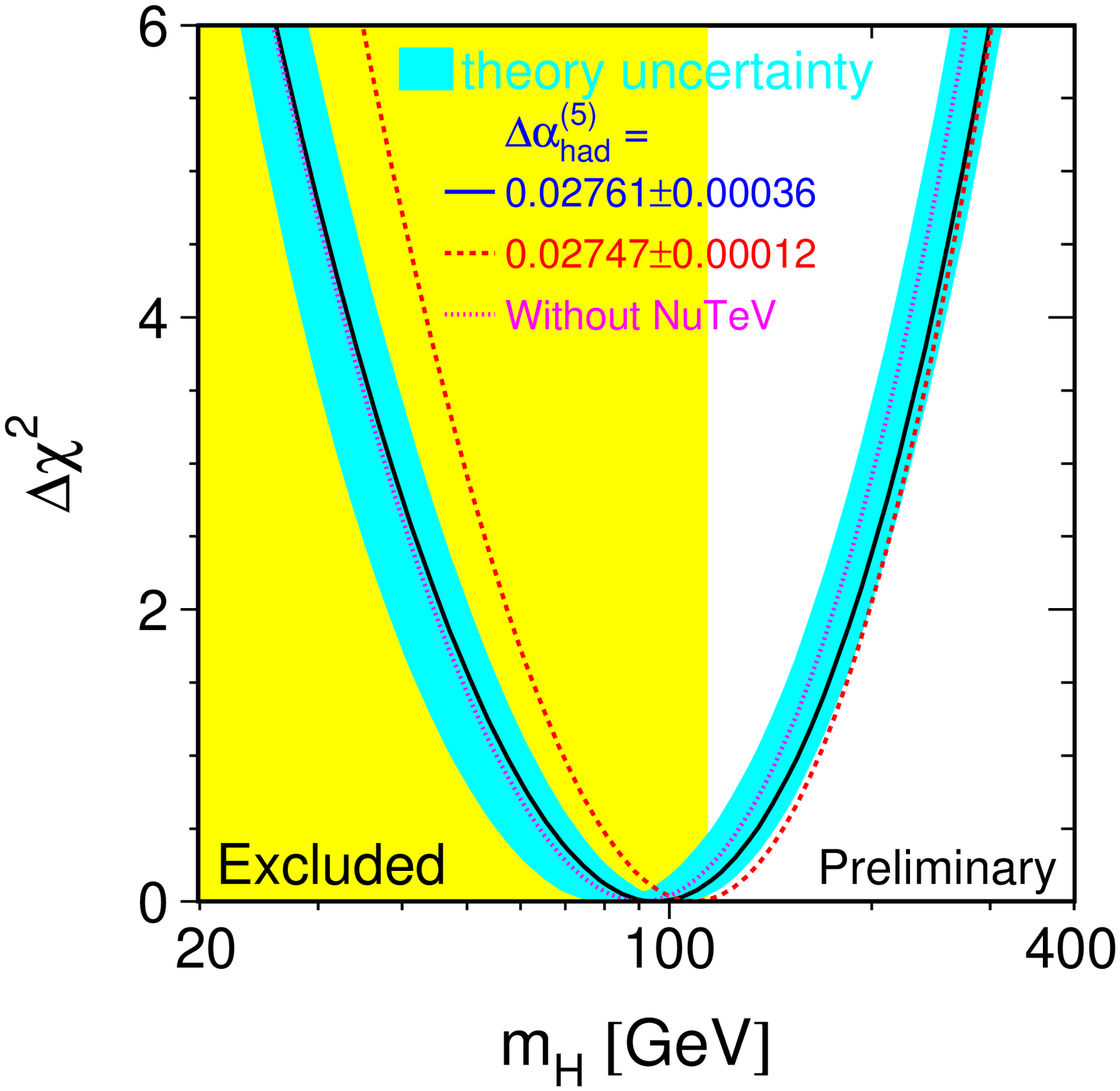,width=4.5cm}
    \vspace{-.7cm}
    \caption{
      World combined electroweak results.
      \label{fig:worldew}
      }
\end{center}
\end{figure}

The combined electroweak data is often summarized as shown
in Figure~\ref{fig:worldew}.
The first plot in this figure shows \Mw\ versus \Mt\ :
the direct measurements, the indirect electroweak data, and
the Standard Model prediction as a function of the Higgs mass.
It can be seen that the precise input data from LEP and SLD
 predicts values of \Mw\
and \Mt\
consistent with those observed, demonstrating that electroweak
correction can correctly predict the mass of heavy particles.
It is observed that both input data and direct measurement of 
\Mw\ and \Mt\ favour a  
light Higgs. 
It can also be seen from this plot that significant improvements
in the uncertainty on \Mw\ will not be very useful if they
are not accompanied by comparable improvements in \Mt.
The second plot shows the variation of the minimum value of the 
$\chi^2$ as a function of $\rm M_H$ for the full electroweak fit.
 The best-fit value
of the Higgs mass is $\rm M_H = 96 ^{+60}_{-38} GeV$, where the error is 
asymmetric as the leading corrections depends on $\rm logM_H$, from 
which the constraint $\rm M_H < 219 GeV$ at 95\% C.L. can be derived.

The next five years will see measurements of similar precision
performed at the Tevatron with the advent of \RunII.
Further substantial improvement in precision will have to wait for 
the Large Hadron Collider and the future Linear Collider. 
 
%
%
\section{Acknowledgements}
The author would like to thank Arno Straessner and Richard Hawkings 
who helped in preparing this talk.
%

%
\end{document}